\documentclass[preprint]{aastex}
\usepackage[english]{babel}
\usepackage{graphics,color} 
\usepackage{epsfig}
\usepackage{natbib}

\newcommand{\lir}{$L_\mathrm{dust} / L_\star$}
\newcommand{\um}{$\mu$m}

\shorttitle{The Exozodiacal Dust Problem}
\shortauthors{Roberge et al.}

\begin{document}

\title{The Exozodiacal Dust Problem for Direct Observations of ExoEarths}

\author{Aki Roberge\altaffilmark{1}, 
Christine H. Chen\altaffilmark{2},
Rafael Millan-Gabet\altaffilmark{3},
Alycia J.~Weinberger\altaffilmark{4},
Philip~M.~Hinz\altaffilmark{5},
Karl R.~Stapelfeldt\altaffilmark{1},
Olivier Absil\altaffilmark{6},
Marc J.~Kuchner\altaffilmark{1}, 
Geoffrey Bryden\altaffilmark{7},
\& the NASA ExoPAG SAG~\#1 Team\altaffilmark{8}}

\email{Aki.Roberge@nasa.gov, cchen@stsci.edu, rafael@ipac.caltech.edu, alycia@dtm.ciw.edu, phinz@as.arizona.edu, Karl.R.Stapelfeldt@nasa.gov,
absil@astro.ulg.ac.be, Marc.Kuchner@nasa.gov, Geoff.Bryden@jpl.nasa.gov}

\altaffiltext{1}{Exoplanets and Stellar Astrophysics Laboratory, Code 667, NASA Goddard Space Flight Center, Greenbelt, MD 20771, USA}
\altaffiltext{2}{Space Telescope Science Institute, 3700 San Martin Drive, Baltimore, MD 21218, USA}
\altaffiltext{3}{California Institute of Technology, NASA Exoplanet Science Institute, Pasadena, CA 91125, USA}
\altaffiltext{4}{Department of Terrestrial Magnetism, Carnegie Institution of Washington, 5241 Broad Branch Road NW, Washington, DC 20015, USA}
\altaffiltext{5}{Steward Observatory, The University of Arizona, 933 North Cherry Avenue, Tucson, AZ 85721, USA}
\altaffiltext{6}{D\'{e}partement d'Astrophysique, G\'{e}ophysique et Oc\'{e}anographie, Universit\'{e} de Li\`{e}ge, 17 all\'{e}e du Six Ao\^{u}t, 4000, Li\`{e}ge, Belgium}
\altaffiltext{7}{Jet Propulsion Laboratory, California Institute of Technology, 4800 Oak Grove Drive, Pasadena, CA 91109, USA}
\altaffiltext{8}{NASA Exoplanet Exploration Program Analysis Group (ExoPAG), 
Debris Disks and Exozodiacal Dust Study Analysis Group (SAG \#1).
Current members: O.~Absil, J.-C.~Augereau, G.~Bryden, J.~H.~Catanzarite, C.~H.~Chen, T.~P.~Greene, P.~M.~Hinz, M.~J. Kuchner, C.~M.~Lisse, B.~A.~Macintosh, R.~Millan-Gabet, M.~C.~Noecker, S.~T.~Ridgway, A.~Roberge (Team Lead), R.~Soummer, K.~R.~Stapelfeldt, C.~C.~Stark, A.~J.~Weinberger, \& M.~C.~Wyatt}

\slugcomment{Accepted for publication in \emph{PASP}, 2012 June 7}

%%%%%%%%%%%%%%%%%%%%%%%%%%%%%%%%%%%%%%%%%%%%%%%%%%%%%%%%%%%%%%%%%%%%%%%%%%%%%%%%

\begin{abstract}

Debris dust in the habitable zones of stars -- otherwise known as exozodiacal dust -- comes from extrasolar asteroids and comets and is thus an expected part of a planetary system.
Background flux from the Solar System's zodiacal dust and the exozodiacal dust in the target system is likely to be the largest source of astrophysical noise in direct observations of terrestrial planets in the habitable zones of nearby stars.
Furthermore, dust structures like clumps, thought to be produced by dynamical interactions with exoplanets, are a possible source of confusion.  
In this paper, we qualitatively assess the primary impact of exozodical dust on high-contrast direct imaging at optical wavelengths, such as would be performed with a coronagraph.
Then we present the sensitivity of previous, current, and near-term facilities to thermal emission from debris dust at all distances from nearby solar-type stars, as well as our current knowledge of dust levels from recent surveys.
Finally, we address the other method of detecting debris dust, through high-contrast imaging in scattered light.
This method is currently far less sensitive than thermal emission observations, but provides high spatial resolution for studying dust structures.
This paper represents the first report of NASA's Exoplanet Exploration Program Analysis Group (ExoPAG).

\end{abstract}


%%%%%%%%%%%%%%%%%%%%%%%%%%%%%%%%%%%%%%%%%%%%%%%%%%%%%%%%%%%%%%%%%%%%%%%%%%%%%%%%

\section{\large Introduction}

Interplanetary dust interior to the Solar System's asteroid belt is called the zodiacal dust, which comes from comet comae and asteroid collisions, just like the dust in any debris disk.
At least 16\% of nearby solar-type stars harbor cold, outer debris dust much denser than the dust in the Solar System's Kuiper belt \citep{Trilling:2008}.
Sensitive new far-IR surveys probing similar stars for cold dust down to about 10 times the Solar System Kuiper belt level find a detection rate of $\sim 25\%$ (Eiroa et al., in preparation).
Unfortunately, we know little about warmer exozodiacal dust (exozodi) in the inner reaches of nearby systems, i.e.\ within the stars' habitable zones.
About 1\% of nearby solar-type stars show a large amount of emission from warm dust \citep[$\gtrsim 1000$ times the expected zodiacal dust emission in the wavelength range from 8.5~\um\ to 12~\um;][]{Lawler:2009}.
However, a more sensitive survey for exozodiacal dust around a smaller set of nearby stars with the Keck Nulling Interferometer (KIN) found mostly non-detections \citep[discussed further below;][]{Millan-Gabet:2011}.
As will be shown, background flux from our local zodiacal dust and the exozodi will likely dominate the signal of an Earth-analog exoplanet (an exoEarth, for short) in direct images and spectra, even if exozodi levels are no greater than the Solar System level.

\begin{figure} \centering 
\plotone{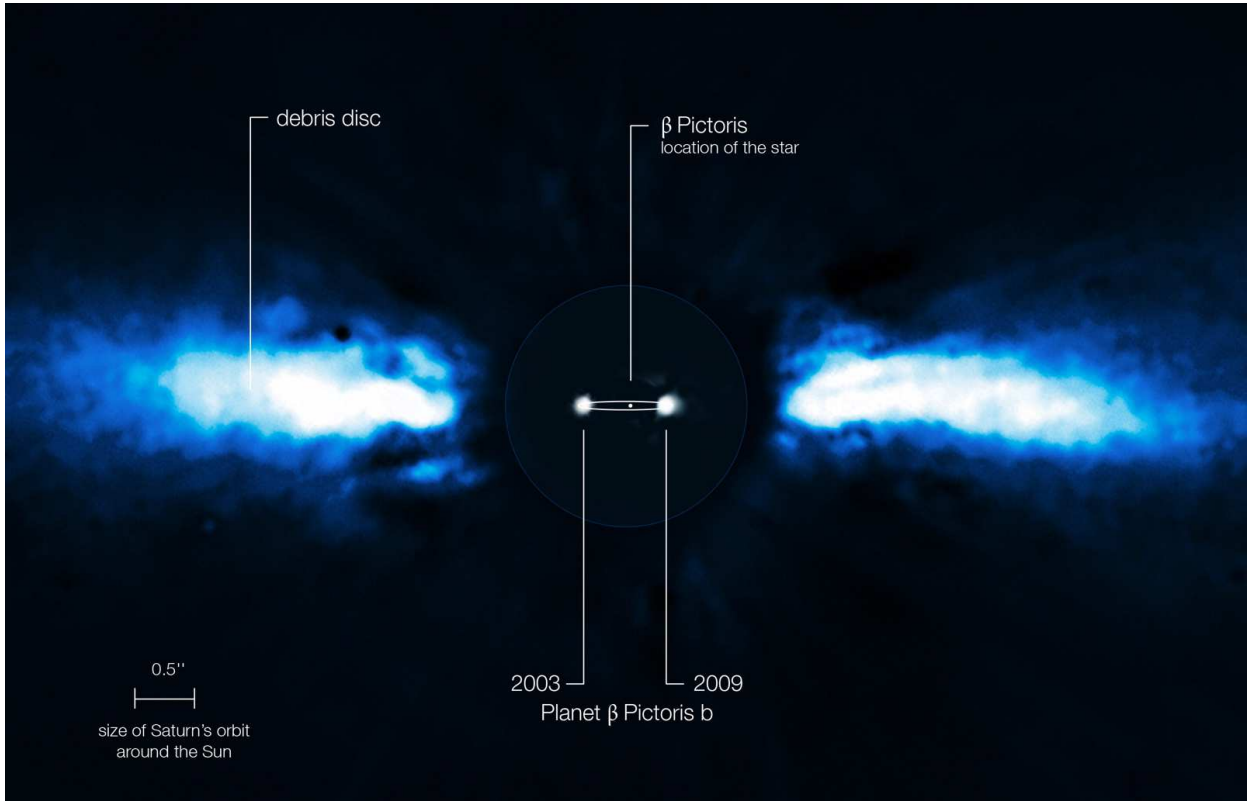}
\caption{A composite image of $\beta$~Pictoris system showing the edge-on debris disk and the exoplanet $\beta$~Pic~b \citep{Lagrange:2010}.
The disk image was obtained with the ESO 3.6-m Telescope; the innermost portion of the disk is obscured.
The planet, imaged at two epochs with ESO's VLT, is likely responsible for a warp seen in the inner disk \citep{Heap:2000}, revealed as an inclined sub-disk in HST ACS coronagraphic images \citep{Golimowski:2006}.
Image credit: ESO$/$A.-M.~Lagrange. \label{fig:betapic}}
\end{figure}

Furthermore, debris disks often display dust structures, such as narrow rings \citep[e.g.\ HR~4796A;][]{Schneider:2009}, clumps \citep[e.g.\ $\epsilon$~Eri;][]{Greaves:2005}, and warps$/$inclined sub-disks \citep[e.g.\ AU~Mic;][]{Krist:2005}. 
A notable example of a warped disk -- the famous $\beta$~Pictoris system -- is shown in Figure~\ref{fig:betapic}.
Such structures are thought to be produced by the dynamical influence of a planet, as are the clumps of zodiacal dust leading and trailing the Earth in its orbit \citep{Dermott:1994}.
Dust clumps are likely to be the most troublesome source of confusion in direct imaging of Earth-size planets, since dynamical clumps orbit the star, though not necessarily with the same period as the perturbing planet 
\citep[e.g.][]{Kuchner:2003}.

Therefore, exozodiacal dust complicates direct imaging of exoplanets in two ways: 1) as a source of noise and 2) as a source of confusion.
The exozodical dust levels around nearby stars will be as important to the success of efforts to find and characterize Earth-like exoplanets as the fraction of stars with potentially habitable planets ($\eta_\mathrm{Earth}$).
There is a strong need to sensitively probe nearby stars for small amounts of dust in their habitable zones.
In this paper, we will first demonstrate the effect of exozodi emission as a source of increased noise in direct imaging of exoplanets, then assess the current knowledge of debris dust abundances and expected progress toward increasing our knowledge.

\subsection{Quantifying Debris Dust: What is a ``Zodi?'' 
\label{sub:zodi}}

Debris disk observers typically express the amount of dust in a debris disk using the system's fractional dust luminosity, \lir, which is the light absorbed by the dust and reemitted at thermal wavelengths (infrared to millimeter) relative to the stellar luminosity. 
This parameter, often also called the fractional infrared luminosity, is determined by integrating the total long-wavelength excess flux seen in a disk spectral energy distribution (SED) over frequency.  
Therefore, it can be measured from unresolved photometry, more easily at long wavelengths where the stellar emission is relatively faint.
\lir\ is not a unit of optical depth, dust mass, or surface brightness. 
In the optically thin case, it is proportional to the dust mass but is affected by grain properties like size and composition.

In the context of direct exoEarth observations, the abundance of debris dust in habitable zones is typically given in units of ``zodis,'' which arose as a convenient way of quickly expressing some sort of ratio to the Solar System zodiacal dust.
However, extreme caution must be taken when interpreting the unit, since different workers have used different dust properties when constructing their one zodi.
The definition adopted largely depended on what was useful or possible for the analysis at hand.
For example, \citet{Gaidos:1999} defined one zodi (spelled ``zody'') as the effective emitting area of the Solar System's zodiacal dust.
\citet{Roberge:2010} took one zodi to be the fractional dust luminosity of the zodiacal dust \citep[\lir~$\sim 10^{-7}$;][]{Nesvorny:2010}, in order to relate the dust abundances of known debris disks to the Solar System (by this standard, $\beta$~Pictoris has about 10,000 zodis of cold dust).

Others assume one zodi is a debris disk that is identical to the Solar System's zodiacal dust in \emph{every respect}, including total mass, spatial distribution, grain size distribution, albedo, scattering phase function, etc.\ \citep[e.g.][]{Millan-Gabet:2011}.
Here, we will refer to such a disk as a zodiacal-twin disk;
around a Sun-twin star, we call it a Solar System-twin disk and it 
has \lir~$\sim 10^{-7}$.
However, care must be taken when applying a zodiacal-twin disk to a star different than the Sun, since such a disk around another type of star may not be both self-consistent and truly identical to the zodiacal dust in every way.
ZODIPIC\footnote{ Available for download at \url{http://asd.gsfc.nasa.gov/Marc.Kuchner/home.html}.} is a publicly available code frequently used to calculate zodiacal-twin disks for any type of star, using the zodiacal dust model in \citet{Kelsall:1998}, which was itself based on fits to observations of the zodiacal dust emission with the COBE DIRBE instrument.
The code does not change the dust density distribution when varying the central star, as expected for a zodiacal-twin disk.
However, the inner disk radius set by the dust sublimation temperature does move outward with increasing stellar luminosity, to prevent an unphysical situation with dust at impossibly high temperatures.
The unfortunate consequence of this is that the total dust mass and \lir\ value of a ZODIPIC zodiacal-twin disk decreases as the stellar luminosity increases.

Nonetheless, for direct observations of exoplanets at any wavelength, it is the surface brightness of the dust thermal or scattered emission near the location of the exoplanet that matters.
While \lir\ can be measured for debris disks, it is not straightforward to convert \lir\ into a surface brightness; one must assume the spatial distribution of the dust, as well as many other properties that are also typically unknown (e.g.\ albedo).
But declaring one zodi to be a zodiacal-twin disk is also perilous.
For most debris disks, we do not know if the inner dust matches the zodiacal dust and in some cases we know it cannot \citep[e.g.\ HD~69830;][]{Beichman:2005}.
Furthermore, this definition is typically applied to stars that are not identical to the Sun without any adjustment to the radial dust distribution, which is not likely to provide an accurate comparison to the Solar System (discussed further below).
Because of these problems, we eschew use of the unit ``zodi'' in this paper, although we acknowledge its convenience and anticipate its continued usage.

\subsection{Exozodi Surface Brightness \label{sub:surf}}

In the next section of this paper, we show the impact of exozodi emission on direct observations of exoEarths at optical wavelengths.
To do this, we need a reference value for the scattered light surface brightness.
For a Solar System-twin disk viewed at $60^\circ$ inclination (the most likely value in a random distribution of inclinations), the modeled surface brightness at a projected separation of 1~AU in the plane of the disk (the quadrature position for a planet) and $\lambda = 0.55$~\um\ is about $22$~mag~arcsec$^{-2}$. 
Other brightness values calculated with ZODIPIC appear in Table 1.B-1 of the TPF-C STDT Final Report\footnote{ \url{http://exep.jpl.nasa.gov/files/exep/STDT\_Report\_Final\_Ex2FF86A.pdf}}.
The inclination of the system strongly affects the observed surface brightness;  at 1~AU, the surface brightness of a ZODIPIC zodiacal-twin disk viewed edge-on is nearly 3~times greater than when viewed face-on.

To apply the above surface brightness value to another star, we take advantage of the fact that by definition the habitable zone is the region around a star where an Earth-like planet receives the right amount of energy for it to have liquid water on its surface.
Therefore, the habitable zone moves with changing stellar luminosity to keep the incident flux constant, such that
\begin{equation}
r_\mathrm{ \: EEID} = 1 \ \mathrm{AU} * \sqrt{ L_{\star} / L_{\sun} } \; ,
\end{equation} 
where EEID stands for ``Earth equivalent insolation distance.''
No matter the star, bodies in the habitable zone always receive about the same total amount of flux; if their physical properties are the same, they also reflect or emit about the same total flux. 

One consequence of this is that the absolute bolometric magnitude of an Earth-twin in the habitable zone does not change from star to star, assuming the same viewing geometry.
Another is that, assuming identical grain properties and inclination, the same amount of dust in the habitable zones of different stars has roughly the same surface brightness.
Therefore, we can reasonably adopt the surface brightness at 1~AU of a Solar System-twin disk viewed at $60^\circ$ inclination as a reference value for the exozodi brightness at the EEID for all stars.

The movement of the habitable zone with stellar luminosity explains the aforementioned problem with applying an unmodified zodiacal-twin disk to stars that are not Sun-like. 
For an earlier-type$/$later-type star, the EEID will be at a larger$/$smaller radius than 1~AU, probing a more$/$less distant region of the zodiacal-twin disk where the surface density may not be the same; in fact, the \citet{Kelsall:1998} model says it should be lower$/$higher. 
If one was expecting the same amount of dust in the star's habitable zone as in the Solar System's, the zodiacal-twin disk would need to be stretched$/$shrunk in radius by the square-root of the stellar luminosity.

%%%%%%%%%%%%%%%%%%%%%%%%%%%%%%%%%%%%%%%%%%%%%%%%%%%%%%%%%%%%%%%%%%%%%%%%%%%%%%%%

\section{\large Impact of Exozodical Emission on Exoplanet Direct Imaging 
\label{sec:impact}}

The importance of zodiacal and exozodiacal background emission is shown by comparing the counts from both sources to the counts from an exoEarth.
Note that the following discussion applies to ordinary high-contrast imaging, such as would be performed with a coronagraph.
The impact of exozodi on mid-IR nulling interferometers intended for direct exoEarth observations (e.g.\ ESA's DARWIN mission concept) is extensively discussed in \citet{Defrere:2010}.
Following the approach in \citet{Brown:2005} and assuming a uniform exozodi distribution near the planet location, the counts ratio is 
\begin{equation} \label{eq:counts}
\frac{C_\mathrm{background}}{C_\mathrm{planet}} = \frac{C_\mathrm{zodi}
+ C_\mathrm{exozodi}}{C_\mathrm{planet}} =
\frac{n_\mathrm{pix} \: \Omega \left( 10^{-0.4 \: z} \:  + \epsilon \: 10^{-0.4 \: x} \right)}{10^{-0.4 (M_\mathrm{p} + 5.0 * \log_{10} d - 5.0)} } \; ,
\end{equation}
where $n_\mathrm{pix}$ is the number of pixels in a critically sampled, diffraction-limited point-spread function (PSF) from a partially obscured circular aperture ($n_\mathrm{pix} = 1 / \mathrm{sharpness} \approx 1 / 0.07 \approx 14.3$); 
$\Omega~=~(\frac{\lambda}{2.0 \, D} \: \frac{206265 \ \mathrm{arcsec}}{1 \ \mathrm{radian}})^2$ is the angular area of a pixel at wavelength $\lambda$ for a telescope of diameter $D$;
$z$ is the surface brightness of the local zodiacal dust, which depends on the direction to the target star (a generic value is $\approx 23$~mag~arcsec$^{-2}$ at $\lambda = 0.55$~\um);
$\epsilon$ is the exozodi surface brightness at the EEID relative to the 
brightness of a Solar System-twin disk;
$x$ is the surface brightness at the EEID (1~AU) of a Solar System-twin disk viewed at $60^\circ$ inclination ($\approx 22$~mag~arcsec$^{-2}$ for $\lambda = 0.55$~\um);
$M_p$ is the absolute magnitude of the planet ($M_V = 29.7$~mag for an Earth-twin orbiting at the EEID, viewed at quadrature);
and $d$ is the distance to the system in pc.

\begin{figure} \centering
\epsscale{1.0}
\plottwo{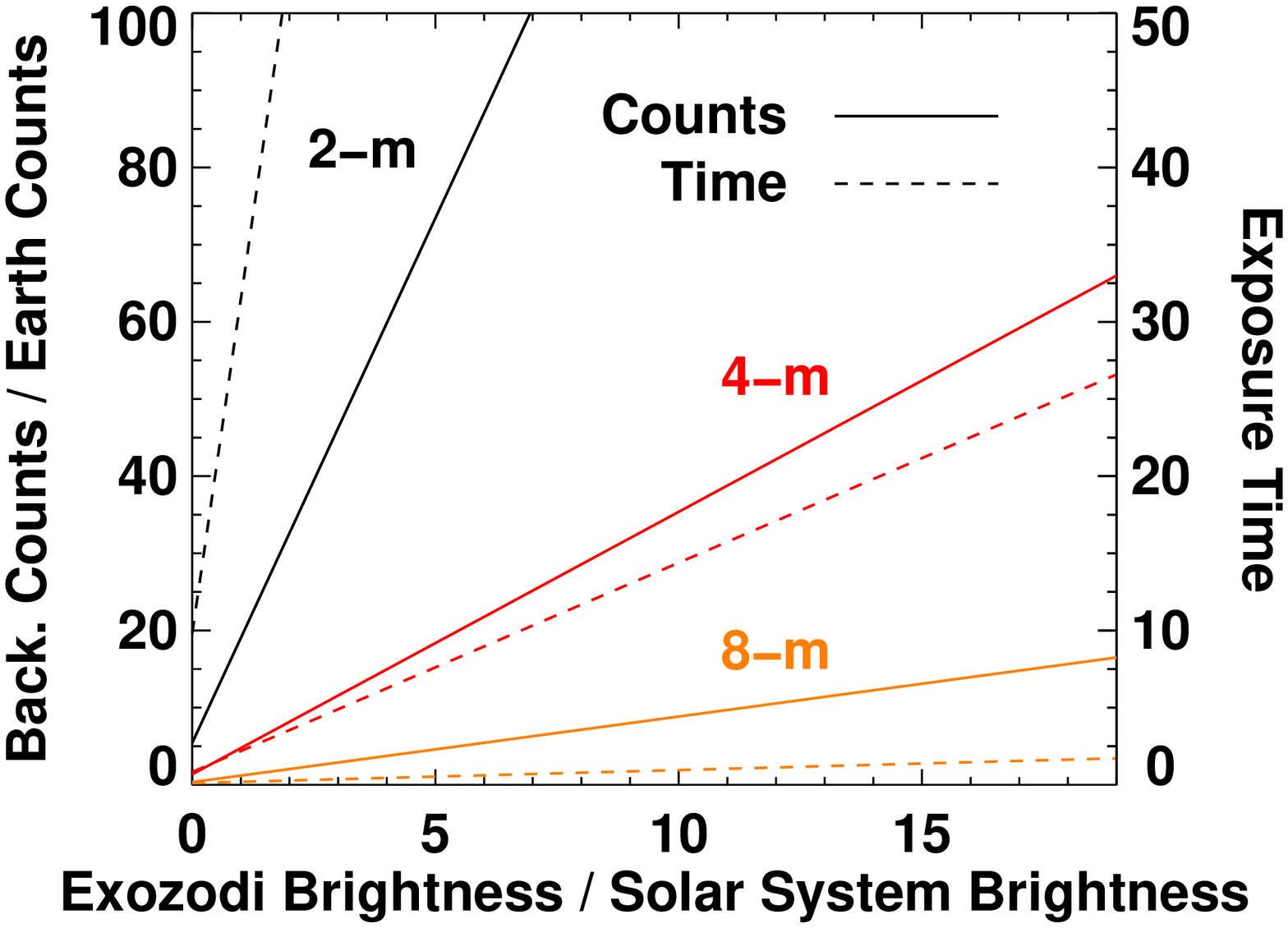}{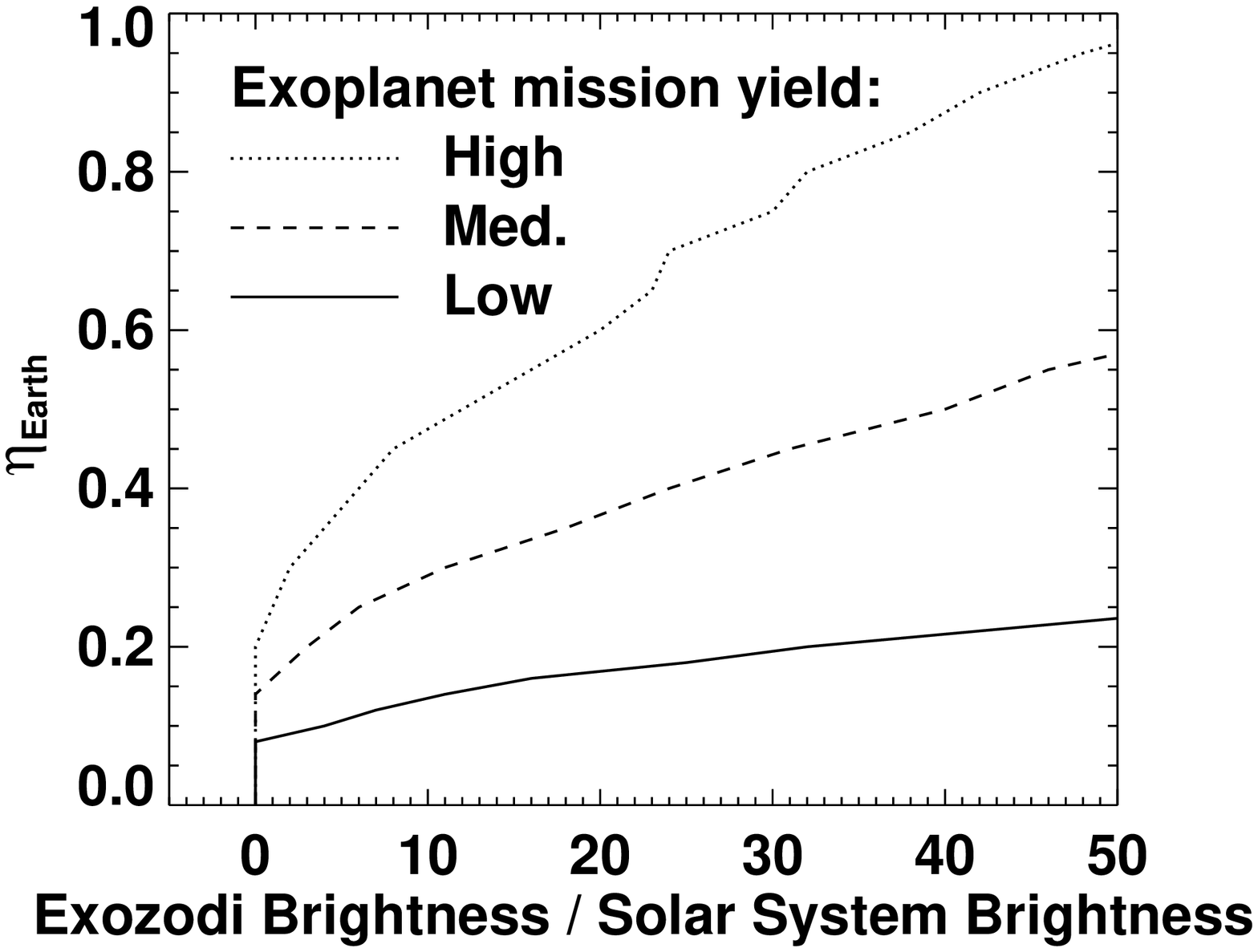}
\caption{Impact of exozodiacal dust emission on direct observations of exoEarths.
For both plots, we assumed the exozodi has a uniform surface brightness. 
Both x-axes give the exozodi surface brightness at the EEID relative to 
the brightness of a Solar System-twin disk ($\epsilon$ in Equations~\ref{eq:counts} \& \ref{eq:time}).
{\bf \ Left:}~Dust counts and imaging exposure time versus exozodi brightness.
The left y-axis is the background counts from zodiacal plus exozodiacal dust divided by the counts from an Earth-twin around a Sun-twin at 10~pc, both calculated for $V$ band (solid lines).
The right y-axis is the exposure time (in arbitrary units) required to detect the Earth-twin with some $S/N$, assuming background-limited imaging (dashed lines).
The counts ratios and exposure times were calculated for three different telescope aperture diameters: 2~meters (black lines), 4~meters (red lines), and 8~meters (orange lines). Smaller apertures are more sensitive to background emission. 
{\bf \ Right:}~The combined effect of exozodi emission and $\eta_\mathrm{Earth}$ on the yield of a direct imaging$/$spectroscopy exoplanet mission.
The y-axis is the fraction of stars with an Earth-size planet in the habitable zone.
The curves were created using a simple mission planning code that chooses real stars within 30~pc for observation until 5~mission years is reached, assuming a 4-meter telescope, some value of $\eta_\mathrm{Earth}$, and that all the stars have the same exozodi level.
At each value of $\eta_\mathrm{Earth}$, the ``Tolerable Exozodi'' (x-axis) is the largest exozodi level for which the desired mission yield (expected number of exoEarths characterized) is achieved.
The analysis was performed for three values of mission yield: high (dotted line), medium (dashed line), and low (solid line).
The smaller $\eta_\mathrm{Earth}$ is, the lower the exozodi level that can be tolerated while still characterizing the desired number of exoEarths.
A full description of the code appears in \citet{Turnbull:2012}; however, the general behavior shown here should be similar for all direct observation exoEarth missions. \label{fig:impact}}
\end{figure}

The left-hand panel of Figure~\ref{fig:impact} shows the counts ratio versus exozodi brightness ($\epsilon$), assuming $\lambda = 0.55$~\um, $d = 10$~pc, the parameter values given above, and three different telescope diameters.
For a 4-meter telescope, the counts from the local zodiacal dust plus a Solar System-twin disk of exozodi viewed at $60^\circ$ inclination are about 5~times greater than the counts from the Earth at 10~pc.
While this may sound disastrous, astronomers commonly observe targets that are fainter than the surrounding background, at the price of increased noise and longer exposure times. 
Assuming background-limited imaging and including unsuppressed light from the central star, the time to detect an exoplanet at some $S/N$ in an image is
\begin{equation}  \label{eq:time} \small
t_\mathrm{image} =  \frac{2 \ n_\mathrm{pix} \ \lambda^2}{\pi \ F_0 \ \Delta \lambda \ D^4 \ T} \left( \frac{S}{N} \right)^2 
10^{0.8 \; (M_p + 5 \log \, d - 5)} \left[ \, A^2 \left( 10^{-0.4 \: z} + \epsilon \ 10^{-0.4 \: x} \right) + \zeta \; 10^{-0.4 \; m_\star } \left( \frac{\pi \; D^2}{4 \; \lambda^2} \right) \right] 
\end{equation}
where $F_0$~is the specific flux for zero magnitude in the image bandpass
($\approx 9500 \ \mathrm{counts} \ \mathrm{s}^{-1} \ \mathrm{cm}^{-2} \
\mathrm{nm}^{-1}$ in $V$ band);
$\Delta \lambda$~is the image bandpass width;
$T$~is the total facility throughput;
$A~=~\left( \frac{206265 \ \mathrm{arcsec}}{1 \ \mathrm{radian}} \right)$;
$\zeta$~is the contrast level at the position of the exoplanet with respect to the theoretical Airy peak of the stellar image ($<10^{-10}$ for instruments designed for direct imaging of exoEarths); and
$m_\star$~is the stellar apparent magnitude
\citep[for more details on the derivation of this equation, see][]{Brown:2005}.

The terms within the square brackets of Equation~\ref{eq:time} account for the background sources: the first for local zodiacal dust, the second for exozodi, and the third for unsuppressed starlight.
The third term is much smaller than the sum of the other two for all known mission concepts designed for direct exoEarth observations at optical$/$near-IR wavelengths.
This equation does not include any dark current or read noise, which are negligible compared to the other noise terms. 
The important points to take away are that in this idealized situation 
1)~the exposure time decreases as $D^4$ -- one factor of $D^2$ for the collecting area and another for the smaller PSF leading to less background blended with the planet -- and 2)~the time increases linearly with increasing exozodi surface brightness.
This is shown in the left-hand panel of Figure~\ref{fig:impact}, where the imaging time is also plotted for the three telescope diameters.

The goal of a mission to obtain direct exoEarths observations is to maximize the number of planets imaged and then characterized with spectra, while maintaining a reasonable mission architecture and lifetime.
The right-hand panel of Figure~\ref{fig:impact} shows how $\eta_\mathrm{Earth}$ and the exozodi level combine to impact the performance of such a mission (ignoring confusion).
The plot was created using a simple mission planning code that chooses real stars within 30~pc of the Sun for observation until the mission lifetime is reached \citep[the mission parameters and planning code are detailed in][]{Turnbull:2012}.
It shows that the smaller $\eta_\mathrm{Earth}$ is, the lower the exozodiacal dust level that can be tolerated while still obtaining direct observations of a sufficient number of exoplanets over the mission lifetime.

The analysis in this section demonstrates the primary effect of exozodi on direct exoEarth observations, i.e.\ increased background noise and longer exposure times.
However, Equation~\ref{eq:time} applies exactly only in the ideal case of uniform background emission and no systematic errors (i.e.\ no speckles).
Furthermore, the methods used to extract the exoplanet signal from the background will affect the $S/N$ achievable.  
If the background contains no unresolved structures, then matched filtering should be an efficient way of identifying point-sources like exoplanets 
\citep[see][]{Kasdin:2006}.
However, since unresolved dust clumps and other background sources (e.g.\ stars and galaxies) should often be present, this technique will not resolve all confusion issues.
Another option is to produce a model for the exozodi emission, fitting to portions of the exoplanet image, then subtract the model from the image.
This should remove smoothly varying exozodi emission fairly easily.
To account for unresolved dust clumps, more complex dynamical modeling of the whole system (dust and exoplanets together) may be required. 
There are also observing strategies to mitigate sources of confusion (e.g.\ multi-epoch imaging, multi-color imaging, or even direct spectroscopy).
But as of this writing, the problem of confusion for direct exoplanet imaging has not been thoroughly studied.

%%%%%%%%%%%%%%%%%%%%%%%%%%%%%%%%%%%%%%%%%%%%%%%%%%%%%%%%%%%%%%%%%%%%%%%%%%%%%%%%

\section{\large Thermal Emission from Dust} 

Now that we have established the importance of exozodi, we assess our knowledge of the debris dust levels around nearby solar-type stars.
Assuming the stellar spectrum is well-described by a Rayleigh-Jeans law (as it typically is at mid-IR and longer wavelengths) and the dust thermal emission is single-temperature blackbody radiation, the fractional dust luminosity may be expressed as 
\begin{equation}
\frac{L_\mathrm{dust}}{L_\star} = \left( \frac{F_\mathrm{dust}}{F_\star} \right) \frac{k \, T_{d}^{4} \left( e^{h \nu / k T_d} - 1 \right)}{h \, \nu \, T_{\star}^{3}} \: , \label{eq:lir}
\end{equation}
where $F_\mathrm{dust}$ and $F_\star$ are the dust and stellar fluxes at some frequency $\nu$, $T_d$ is the dust temperature, and $T_\star$ is the stellar effective temperature \citep[e.g.][]{Bryden:2006}.
For blackbody grains, assuming all the dust is at the same temperature is equivalent to assuming that all the grains are at the same distance from the star (i.e.\ a ring-like disk).
In reality, grains of different sizes will have different temperatures at the same distance.
That being said, most debris disk SEDs are fairly well-described by either 
a cold ($\sim$~tens of~K) single-temperature blackbody -- an outer dust ring -- or the sum of a cold blackbody and a warmer one -- outer and inner dust rings, an architecture reminiscent of the Solar System's Kuiper and asteroid belts \citep[e.g.][]{Chen:2006,Chen:2009}.

\begin{figure} \centering 
\plotone{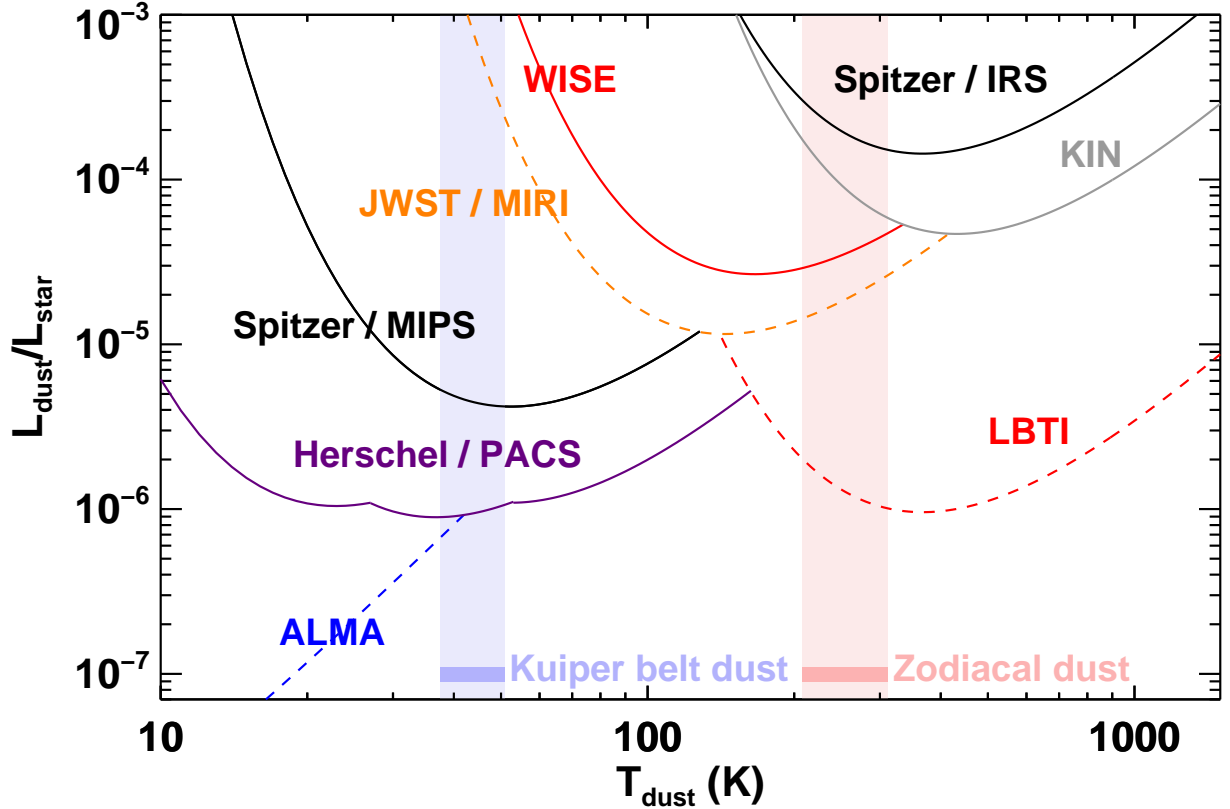}
\caption{Sensitivity limits for detection of debris dust around nearby Sun-like stars, for various recent (Spitzer, WISE), current (KIN, Herschel) and near-term facilities (LBTI, JWST, ALMA).  
The curves show $3 \sigma$ detection limits in terms of the fractional dust luminosity (\lir) versus its temperature.  
Recent and current facilities are plotted with solid lines, near-term ones with dashed lines.
%On the right y-axis, \lir\ is given in units of zodis.  
The instrument data assumed for these curves appear in 
Table~\ref{tab:sens_data}. 
The temperature ranges for dust in two zones around the Sun are shown with vertical bars, calculated assuming blackbody grains.
The habitable zone (0.8~AU -- 1.8~AU) is shown in pink, the Kuiper belt (30~AU -- 55~AU) is shown in light blue. 
The modeled \lir\ values for the Solar System's Kuiper belt dust 
\citep[$\sim 10^{-7}$;][]{Vitense:2012} and the zodiacal dust 
\citep[$\sim 10^{-7}$;][]{Nesvorny:2010} are marked with horizontal light blue and pink bars. \label{fig:sens_curves}}
\end{figure}

Using Equation~\ref{eq:lir}, we may estimate the sensitivity of recent, current, and near-term telescope facilities to exozodiacal dust at different temperatures.
Figure~\ref{fig:sens_curves} shows the estimated $3 \sigma$ sensitivity curves for the Spitzer Space Telescope (Spitzer), the Wide-field Infrared Survey Explorer (WISE), the Keck Interferometer Nuller (KIN), the Herschel Space Observatory (Herschel), the Large Binocular Telescope Interferometer (LBTI), the James Webb Space Telescope (JWST), and the Atacama Large Millimeter Array (ALMA).
The relevant facility performance parameters -- the instruments, years of operation (actual or expected), observation wavelengths, and uncertainties -- appear in Table~\ref{tab:sens_data}.

To calculate the sensitivity curves, we assumed a stellar effective temperature equal to the Sun's.
For each instrument, the $1 \sigma$ photometric uncertainty at the observation wavelength (Column~5 of Table~\ref{tab:sens_data}) was converted into an uncertainty on the flux relative to the stellar flux at that wavelength 
($\sigma_{F_\mathrm{dust}/F_\star}$ in Column~6).
We then replaced $F_\mathrm{dust} / F_\star$ in Equation~\ref{eq:lir} with three times the relative flux uncertainties and plotted $L_\mathrm{dust} / L_\star$ as a function of $T_\mathrm{dust}$.
Each curve has a minimum at the temperature for which the dust blackbody emission curve peaks at the frequency of observation 
($T_\mathrm{peak}$ in Column~4). 
Emission from dust warmer$/$cooler than $T_\mathrm{peak}$ can be detected with reduced sensitivity.
However, in most systems, emission from dust at temperatures near 
$T_\mathrm{peak}$ will swamp any emission from much warmer$/$cooler dust, which is why a survey for habitable zone dust is best done with an instrument that operates near 10~\um.
Note that although JWST$/$MIRI will have only moderate absolute photometric accuracy, the telescope's large collecting area will allow MIRI to achieve this accuracy for fainter stars than other facilities covering its bandpass. 

The temperature ranges for dust in the habitable zone and in the Kuiper belt of a Sun-like star are overlaid in Figure~\ref{fig:sens_curves}.
The temperature range for the habitable zone is \mbox{208 -- 312~K}, calculated assuming the zone extends from 0.8 AU to 1.8 AU \citep{Kasting:2009}.
For the Kuiper belt, we assumed an annulus extending from 30~AU to 55~AU, giving a temperature range of \mbox{38 -- 51~K.}
In both cases, the dust temperatures were calculated assuming blackbody dust grains in radiative equilibrium with a Sun-like star.
The modeled \lir\ values for the zodiacal dust \citep[$\sim 10^{-7}$;][]{Nesvorny:2010} and dust in Solar System's Kuiper belt \citep[$\sim 10^{-7}$;][]{Vitense:2012} are marked with horizontal bars.
The latter value is highly dependent on poorly known characteristics of the source bodies (Kuiper belt objects); 
it was revised downward by about a factor of 10 in the last two years, based on new knowledge from in-situ detections of dust grains with the Student Dust Counter on-board NASA's New Horizons mission \citep{Poppe:2010,Han:2011}.

For nulling instruments (KIN and LBTI), to convert the null leakage to a relative flux ($F_\mathrm{dust}/F_\star$), one must account for the fact that some dust emission may be removed by the dark fringes, depending on the exact spatial distribution of the dust.  
A recent KIN exozodiacal dust survey found an average value of 0.4 for the fraction of dust emission transmitted through the fringes, assuming unmodified zodiacal-twin disks \citep{Millan-Gabet:2011}.  
The null leakage error must be divided by this transmission factor to give the correct uncertainty on the relative flux.  
However, the correct factor assuming a ring-like dust distribution, as we have done here, is not presently known and could be as large as 1. 
Therefore, we have chosen to ignore the transmission factor in calculating the sensitivity curves for nulling instruments.

%The sensitivity curve for CHARA, which detects circumstellar dust through high %precision visibility measurements using near-IR interferometry, was calculated %using Equation~\ref{eq:lir} in the same fashion as the other curves.  However, %at the short wavelength of these observations (2.2~\um), the spectrum of a G-%type star is not well-described by a Rayleigh-Jeans law.  Assuming a Solar %System-like dust distribution, the actual $3 \sigma$ sensitivity of CHARA to %hot dust is $\sim 9000$~zodis, a factor of about 2 better than implied by %Figure~\ref{fig:sens_curves} (O.\ Absil, personal communication).

Over the last several years, high precision visibility measurements with near-IR interferometers (VLTI$/$VINCI, CHARA$/$FLUOR) have been used to detect circumstellar dust around a few nearby stars \citep[e.g.][]{diFolco:2007}.
These observations can detect large amounts of hot dust (CHARA$/$FLUOR 
$3 \sigma$ sensitivity $\sim 5000 \times$ a Solar System-twin disk for $\sim 1700$~K dust; O.\ Absil, personal communication).
While these observations are most sensitive to dust interior to habitable zones, they raise interesting questions about the origin of large amounts of dust so close to the central stars; one possibility is star-grazing comets \citep{Absil:2006}.
Another new instrument with sensitivity to hot dust is the Palomar Fiber Nuller \citep[PFN;][]{Hanot:2011}.
Recent PFN observations at 2.2~\um\ of a debris disk system with a hot dust detection from CHARA (Vega) limit the dust to $\lesssim 0.2$~AU from the central star \citep[or $\gtrsim 2$~AU if there is a significant scattered light contribution;][]{Mennesson:2011}.
Unfortunately, we cannot accurately estimate the sensitivity of these near-IR instruments with Equation~\ref{eq:lir}, since the assumption that the stellar spectrum is well-described by a Rayleigh-Jeans law breaks down; therefore, we have not included CHARA$/$FLUOR or PFN sensitivity curves in Figure~\ref{fig:sens_curves}.

Examination of Figure~\ref{fig:sens_curves} shows that previous and current facilities (Spitzer, WISE, KIN, Herschel) are relatively insensitive to dust in the habitable zone of Sun-like stars.
To date, the most sensitive survey for habitable zone dust around nearby solar-type stars was performed with the KIN \citep{Millan-Gabet:2011}.
While the mean $3 \sigma$ detection limit for an individual star in this survey was $480 \times$ a zodiacal-twin disk (shown in Figure~\ref{fig:sens_curves}), the ensemble of non-detections was used to infer a $3 \sigma$ upper limit on the mean exozodi level of $150 \times$ a zodiacal-twin disk.
Unfortunately, these limiting values are far too high to reliably estimate the performance of a future mission aimed at direct observations of Earth-analog planets.
The sole near-term facility sensitive enough to assess the habitable zone dust around solar-type stars is LBTI, which should have a $3 \sigma$ detection limit of about $10 \times$ a Solar System-twin disk for $\sim 300$~K dust. 
An LBTI survey for exozodiacal dust around a set of about 60 -- 100 nearby stars is expected to begin in late 2012\footnote{ \url{http://nexsci.caltech.edu/missions/LBTI/cfp\_keysci.shtml}}.
%Observations of additional targets may be required to fully characterize the %typical habitable zone dust levels for nearby solar-type stars, although the %dust sensitivity will decrease for more distant stars.

Turning to dust structures, observations of dust thermal emission are typically unresolved; at best, SED modeling provides information only on the radial distribution of the dust.
Dust clumps caused by an Earth-mass planet orbiting 1~AU from a Sun-like star may be expected to have sizes of about 0.2~AU, corresponding to 0.02~arcsec for a star at 10~pc \citep{Dermott:1994}.
For each facility considered above, the spatial resolution at the reference wavelength is given in Table~\ref{tab:sens_data}.
Of the current and near-term facilities, only ALMA will have spatial resolution high enough to resolve dust clumps as small as those produced by an Earth-mass planet; unfortunately, ALMA is not sensitive to warm dust in habitable zones.
However, note that more massive planets can create larger clumps \citep{Stark:2008}.

Looking beyond the timeframe covered above, the next generation of ground-based Extremely Large Telescopes (ELTs) should begin operation in the early 2020's. 
For these $> 25$-meter apertures, detecting and spatially resolving thermal emission from habitable zone dust around nearby stars should be a tractable problem. 
Assuming a Solar System-twin disk, the exozodi flux at 10~\um\ relative to the stellar flux is $F_\mathrm{dust} / F_\star \approx 5 \times 10^{-5}$, or about $85~\mu$Jy for a system at 10~pc;
this is well within the photometric detection limits of planned ELTs (A.\ Weinberger, personal communication).
The spatial resolution of a 25-meter aperture at that wavelength ($\lambda/D \approx 83$~milliarcsec) could allow us to resolve thermal emission from regions at radii $\gtrsim 0.8$~AU for a system at 10~pc, as long as light from the central star is sufficiently suppressed and removed at such a small angular separation.

\begin{figure} \centering
\plotone{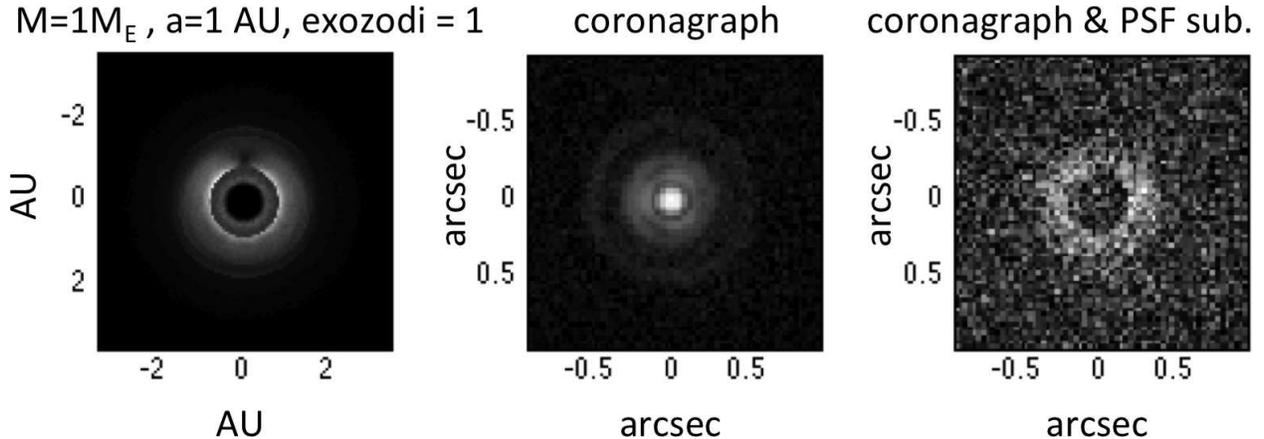}
\caption{Simulated high-contrast imaging of exozodi thermal emission with a ground-based ELT.
{\bf Left:}~A model for a face-on debris disk around $\tau$~Ceti, a Sun-like star at 3.7~pc \citep{Stark:2008}.
The disk has a dust abundance equal to that of a zodiacal-twin disk and an embedded 1~Earth-mass planet orbiting at 1~AU (planet located at 12 o'clock). Image credit: C.~Stark (model available for download at http://asd.gsfc.nasa.gov/Christopher.Stark/catalog.php).
{\bf Middle:}~A simulated 10~\um\ imaging observation of the model using an idealized coronagraph designed to produce high Strehl ratios ($\sim 98\%$ at 10~\um) on the Giant Magellan Telescope (25-meter aperture); 
theoretically, the instrument suppresses the light from the central star by a factor of $\sim 10^{3}$. 
Telescope and sky background emission were included in the simulation.
{\bf Right:}~The simulated image after subtraction of a reference PSF image,
achieving a contrast of $\sim 10^{-6}$ at an angular separation of $1 \: \lambda / D$ (83~milliarcsec).
Slightly different seeing was assumed for the target and PSF observations.
\label{fig:gmt} }
\end{figure}

So instead of an unresolved measurement of the combined light from the star and dust (all the instruments discussed above do this, except the infrared interferometers and ALMA in the extended array mode), an ELT with advanced high-contrast imaging instrumentation could directly image the thermal emission from habitable zone dust around nearby stars.
In the next few years, new coronagraphs operating at near-IR wavelengths on ground-based telescopes should reach $10^{-6}$ contrast at small angular separations, using extreme adaptive optics and reference PSF subtraction (fuller discussion of coronagraphic imaging appears in the next section).  
This contrast will also be possible at 10~\um\ in the next decade; 
in fact, it will probably be easier to achieve, thanks to the higher Strehl ratio possible at the longer wavelength.   
%
%If this is the case, an ELT could detect the dust near 1~AU in a Solar System-%twin disk at 10~pc in one night (A.~Weinberger, personal communication).
%
A simulation of coronagraphic imaging of habitable zone dust around a nearby solar-type star with an ELT appears in Figure~\ref{fig:gmt}.
It seems likely that by sometime in the 2020s, ELTs will measure low levels of dust in the habitable zones of stars within $\sim 10$~pc.

%%%%%%%%%%%%%%%%%%%%%%%%%%%%%%%%%%%%%%%%%%%%%%%%%%%%%%%%%%%%%%%%%%%%%%%%%%%%%%%%

\section{High-Contrast Imaging in Scattered Light}

\begin{figure} \centering 
\epsfig{file = 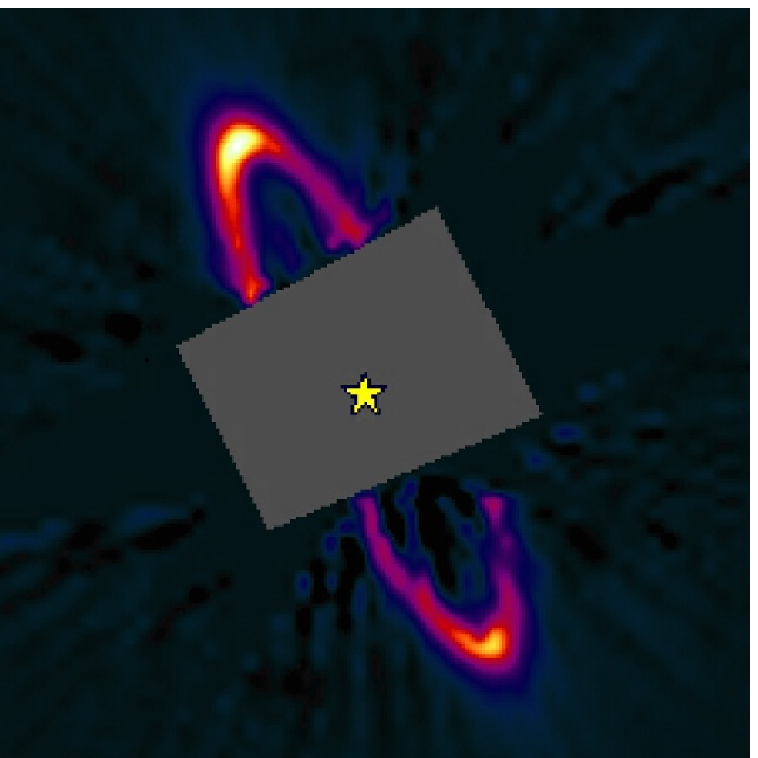, height=3.25in} \hspace*{0.25in}
\epsfig{file = 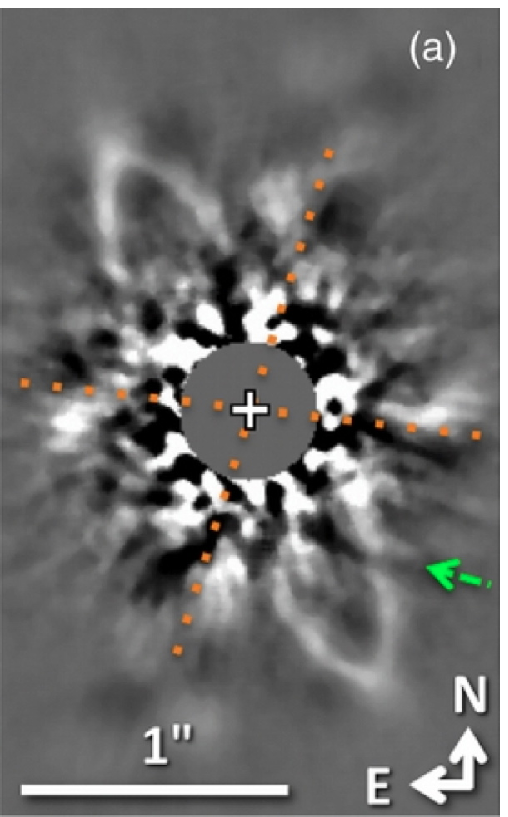, height=3.25in}
\caption{High-contrast imaging of the HR~4796A debris disk in scattered light.
For this disk, \lir~$= 4.7 \times 10^{-3}$ \citep{Moor:2006}. 
{\bf Left: } Optical wavelength HST$/$STIS coronagraphic image.
Image credit: \citet{Schneider:2009}.
{\bf Right: } Near-IR wavelength Subaru$/$HiCIAO image. Image credit: \citet{Thalmann:2011}.
\label{fig:hr4796A} }
\end{figure}

Debris disks may also be detected through coronagraphic imaging of scattered light; such observations have high spatial resolution and are best for revealing dust structures (see Figure~\ref{fig:hr4796A}).
However, current coronagraphy is far less sensitive at all distances from the central stars than observations of thermal emission, and cannot image the habitable zones at all.
This is due to the difficulty of suppressing diffracted and scattered light from the bright central star (using a coronagraph), and then removing residual starlight through subtraction of a reference PSF.
The accuracy with which this can be done largely depends on the stability of the telescope PSF, whether one is using a space-based or ground-based facility.

PSF subtraction can be improved using differential imaging techniques, which distinguish between PSF artifacts and real sources.
Some examples are angular differential imaging (ADI), which images the target at many rotation angles \citep[e.g.][]{Marois:2006}; 
chromatic differential imaging (CDI), which uses the known wavelength dependence of PSF artifacts to identify and remove them \citep[e.g.][]{Crepp:2011};
and polarimetric differential imaging (PDI), which compares images of the target in different polarizations to remove unpolarized starlight \citep[e.g.][]{Quanz:2011}.
Note that ADI and CDI are best at searching for faint point-sources and work less well on extended structures like debris disks.
There are also post-observation techniques that can help with PSF subtraction, like locally optimized combination of images \citep[LOCI;][]{Lafreniere:2007}.

To date, the most sensitive scattered light images of debris disks were obtained with HST, due to its excellent PSF stability compared to ground-based telescopes, even those with adaptive optics (AO) systems.
Several HST instruments have had high-contrast imaging capability;
they are listed in Table~\ref{tab:contrast}.
This table also lists the \lir\ value for the faintest disk successfully imaged in scattered light with each HST instrument.
Examination of these values quickly demonstrates the much greater sensitivity of thermal emission observations (compare to Figure~\ref{fig:sens_curves}).

A new generation of coronagraphs behind extreme AO systems on 8-meter-class ground-based telescopes is in the development and commissioning phase.
These include the High Contrast Instrument for the Subaru Next Generation Adaptive Optics (HiCIAO) instrument on the Subaru Telescope, the Spectro-Polarimetric High-contrast Exoplanet REsearch (SPHERE) instrument on the VLT, and the Gemini Planet Imager (GPI) on the Gemini South Telescope.
HiCIAO has begun science operations \citep[e.g.][]{Thalmann:2011}, while we expect the other two instruments to be commissioned in the next year.
Further in the future, all three imaging instruments planned for JWST (NIRCam, NIRISS, and MIRI) include high-contrast capabilities; NIRCam and NIRISS (formerly known as TFI) will operate in the visible$/$near-IR and be sensitive to scattered light. 
More information about all these instruments and others may be found in \citet{Beichman:2010}.

Due to the inherent difficulties in assessing instrument performance before commissioning and also to the variety of observing and post-observation processing techniques available, it is difficult to predict the ultimate performance of these new instruments for high-contrast imaging of extended structures like debris disks. 
A preliminary measurement of the point-source contrast achievable using HiCIAO with ADI is available (see Table~\ref{tab:contrast}); so far, the faintest disk successfully imaged is HR~4796A, one of the very brightest debris disks (shown in the right-hand panel of Figure~\ref{fig:hr4796A}).
Table~\ref{tab:contrast} also includes predictions for the contrast that will be achievable with GPI and JWST$/$NIRCam; however, these assumed ADI would be used and may not accurately predict the instrument contrast for extended sources.
A recent prediction of the GPI performance for debris disks suggests that compact ($1\arcsec$~diameter) disks with \lir~$\gtrsim 2 \times 10^{-5}$ can be imaged, using PDI and assuming moderately polarizing grains (B.~Macintosh, personal communication).
Unlike the other instruments, JWST$/$NIRISS uses sparse-aperture interferometric imaging to perform high-contrast imaging; a prediction for its point-source contrast appears in Table~\ref{tab:contrast}, but the expected extended-source performance is not yet available. 

%%%%%%%%%%%%%%%%%%%%%%%%%%%%%%%%%%%%%%%%%%%%%%%%%%%%%%%%%%%%%%%%%%%%%%%%%%%%%%%%

\section{Summary}

\begin{enumerate}
\item Exozodiacal dust affects future efforts to directly observe Earth-like planets in the habitable zones of nearby stars in two ways: 1) background flux leading to increased noise and 2) dust structures causing confusion with unresolved exoplanets.
The impact of second problem has not been thoroughly studied.
\item Assuming uniform exozodi surface brightness, a 4-meter telescope aperture, and optical observing wavelengths, the counts from the local zodiacal dust plus a Solar System-twin disk of exozodi are about 5~times greater than the counts from the Earth observed at quadrature orbiting 1~AU from a Sun-like star at 10~pc.
\item LBTI is the only previous, current, or near-term facility sensitive enough to detect exozodiacal dust in the habitable zones of nearby solar-type stars at levels approaching the Solar System zodiacal dust level.
This facility should begin surveying nearby stars for warm dust in 2012.
A decade or so in the future, ground-based ELTs may be able to image low levels of habitable zone dust around stars within $~\sim 10$~pc.
\item Dust structures like clumps located far from the central stars may currently be detected with high-contrast imaging of light scattered from dust disks. 
These observations are currently far less sensitive than observations of unresolved thermal emission.
Cold clumps at large distances may soon also be imaged with ALMA.
Unfortunately, dust structures in habitable zones are likely to prove elusive, although there is a chance that a new generation of ground-based coronagraphs may provide some information for the nearest stars.
\end{enumerate}

%%%%%%%%%%%%%%%%%%%%%%%%%%%%%%%%%%%%%%%%%%%%%%%%%%%%%%%%%%%%%%%%%%%%%%%%%%%%%%%%
%
% ACKNOWLEDGEMENTS
%

\acknowledgements

This work was performed by members of the Debris Disks \& Exozodiacal Dust Study Analysis Group (SAG~\#1), part of NASA's Exoplanet Exploration Program Analysis Group (ExoPAG).

%%%%%%%%%%%%%%%%%%%%%%%%%%%%%%%%%%%%%%%%%%%%%%%%%%%%%%%%%%%%%%%%%%%%%%%%%%%%%%%%
%
% REFERENCES
%
%\bibliographystyle{apj}
%\bibliography{ms_v2}

%%%%%%%%%%%%%%%%%%%%%%%%%%%%%%%%%%%%%%%%%%%%%%%%%%%%%%%%%%%%%%%%%%%%%%%%%%%%%%%%
%
% TABLES
%

\begin{deluxetable}{llccccccc}
\tablewidth{0pt}
\tablecolumns{8}
\tablecaption{Detection of dust thermal emission: data for various telescope facilities \label{tab:sens_data}}
\tablehead{
\colhead{Telescope $/$} & \colhead{Operation} & 
\colhead{$\lambda_\mathrm{obs.}$} & 
\colhead{$T_\mathrm{peak}$\tablenotemark{a}} & 
\colhead{Resolution\tablenotemark{b}} & 
\colhead{Uncertainty\tablenotemark{c}} & 
\colhead{$\sigma_{F_\mathrm{dust}/F_\star}$\tablenotemark{d}} & 
\colhead{Ref.} \\
\colhead{Instrument} & \colhead{Dates} & \colhead{($\mu$m)} & 
\colhead{(K)} & \colhead{($ \, \arcsec \,$) $^{ \ }$} & \colhead{($1 \sigma$)} 
& & }
\startdata
KIN	& 2005 -- 12   & 8.5 & 432 & 0.01\tablenotemark{e} & 0.3\% leak error\tablenotemark{f} & 0.003 & 1 \\
Spitzer$/$IRS   & 2003 -- 09 & 10 & 367 & 2.4 & 1.5\% of star flux & 0.015 & 2\\
LBTI           	& 2012 --    & 10 & 367 & 0.05\tablenotemark{e} & 0.01\% leak error\tablenotemark{f} & 0.0001 & 3 \\
WISE$/$W4	& 2009 -- 11 & 22.1 & 166 & 12 & 3\% of star flux\tablenotemark{g} & 0.03 & 4 \\
JWST$/$MIRI     & 2018 -- 23 & 25.5  & 144 & 0.9 & 2\% of star flux & 0.02 & 5 \\        
Spitzer$/$MIPS  & 2003 -- 09 & 70 & 52 & 18.0 & 15\% of star flux & 0.15 & 6 \\
Herschel$/$PACS & 2009 -- 13 & 70 & 52 & 5.2 &  1.61 mJy & 0.04\tablenotemark{h} & 7 \\
	        &  & 100 & 37 & 7.7 & 1.90 mJy & 0.1\tablenotemark{h} & 7 \\
	        &  & 160 & 23 & 12.0 & 3.61 mJy & 0.4\tablenotemark{h} & 7 \\
ALMA\tablenotemark{i}   & 2012 --    & 1250 & 3 & 0.02 & 0.1 mJy & 0.7\tablenotemark{h} & 8 \\
\enddata
\tablenotetext{a}{Temperature for which the dust blackbody emission peaks at the observation frequency.  Instrument is most sensitive to dust near this temperature.}
\tablenotetext{b}{FWHM of the instrument PSF at the observation wavelength.}
\tablenotetext{c}{Photometric uncertainty at the observation wavelength.} 
\tablenotetext{d}{Photometric uncertainty relative to the stellar flux at the observation wavelength.}
\tablenotetext{e}{$1/2$ the fringe spacing. Spatial resolution only in the direction perpendicular to the fringe pattern, so not applicable for dust clumps.}
\tablenotetext{f}{Null leakage error.}
\tablenotetext{g}{Preliminary value, applicable only for the best cases (D.\ Padgett, personal communication).}
\tablenotetext{h}{Calculated assuming the photometric uncertainty in the previous column and $F_\star$ at the observation wavelength for a Sun-like star at 10~pc.}
\tablenotetext{i}{The ALMA spatial resolution is for the extended array mode, while the photometric uncertainty assumes an unresolved observation with the compact array.}
\tablerefs{
[1]~\citet{Millan-Gabet:2011}; 
[2]~\citet{Beichman:2006a};
[3]~\citet{Hinz:2008};
[4]~User's Guide to the WISE Preliminary Data Release, \url{http://wise2.ipac.caltech.edu/docs/release/prelim/expsup/wise\_prelrel\_toc.html};
[5]~Estimated final absolute photometric accuracy (C.\ Chen, personal communication);
[6]~\citet{Bryden:2006};
[7]~Sensitivity of PACS scan map with on-source integration time $\approx 360$~sec, calculated with HSPOT v6.0.1; 
[8]~Expected full array sensitivity in Band~6 with integration time of 60~sec, ALMA Early Science Primer, v2.2 (May 2011), \url{http://almatelescope.ca/ALMA-ESPrimer.pdf}.
}
\end{deluxetable}

\begin{deluxetable}{llccccc}
\tablewidth{0pt}
\tablecolumns{7}
\tablecaption{High-contrast optical$/$near-infrared imaging of dust scattered light: instrument performance \label{tab:contrast}}
\tablehead{
\colhead{Facility $/$} & \colhead{Operation} & \colhead{IWA\tablenotemark{a}} & \colhead{Contrast\tablenotemark{b}}  & \multicolumn{2}{c}{Faintest Disk Imaged} & \colhead{Refs.}  \\
\colhead{Instrument}  & \colhead{Dates}     & \colhead{($\arcsec$)} & \colhead{at $1\arcsec$} & \colhead{ID} &  \colhead{(\lir)} & \colhead{ } 
}
\startdata
HST$/$STIS  & 1997 -- 2004, 2009 -- & 0.5 & $3 \times 10^{-3} \hspace*{1.ex} ^{ \ }$ & HD202628 & $1 \times 10^{-4}$ & 1, 2 \\
HST$/$NICMOS & 1997 -- 99, 2002 -- 08 & 0.5 & $10^{-5} \hspace*{1.ex} ^{ \ }$ & HD181327 & $2 \times 10^{-3}$ & 3, 4     \\
HST$/$ACS & 2002 -- 07 & 1 & $10^{-5} \hspace*{1.ex} ^{ \ }$ & Fomalhaut & $8 \times 10^{-5}$ & 5, 6 \\
Subaru$/$HiCIAO & 2010 --  & 0.15 & $10^{-4.8}$ \ \tablenotemark{c} & HR4796A & $5 \times 10^{-3}$ & 7, 8 \\
Gemini S$/$GPI & 2012 -- & 0.08 & $\sim 10^{-6 \ \mathrm{to} \ -7}$ \ \tablenotemark{c} & \nodata & \nodata & 9 \\
JWST$/$NIRCam & 2018 -- 23 & 0.3 & $\sim 10^{-5}$ \ \tablenotemark{c} & \nodata & \nodata & 10 \\
JWST$/$NIRISS & 2018 -- 23 & 0.1 & $\sim 10^{-4 \ \mathrm{to} \ -5}$ \ \tablenotemark{c} & \nodata & \nodata & 11 \\
\enddata
\tablenotetext{a}{Inner working angle (smallest achievable).}
\tablenotetext{b}{Relative to peak of unobscured PSF, with reference PSF subtracted.}
\tablenotetext{c}{Assuming a point-source. Will probably be worse for extended sources like disks.} 
\tablerefs{ 
[1]~STIS Instrument Handbook, v10.0, \url{http://www.stsci.edu/hst/stis/documents/handbooks/currentIHB/cover.html}; 
[2]~\citet{Krist:2012};
[3]~\citet{Schneider:2007}; 
[4] \citet{Schneider:2006};
[5]~ACS Instrument Handbook, v10.0, \url{http://www.stsci.edu/hst/acs/documents/handbooks/cycle19/cover.html}; 
[6]~\citet{Kalas:2005}; 
[7]~\citet{Suzuki:2010};
[8]~\citet{Thalmann:2011};
[9]~GPI web page, \url{http://planetimager.org/pages/gpi\_tech\_contrast.html};
[10]~\citet{Krist:2007},
[11]~STScI NIRISS web page, \url{http://www.stsci.edu/jwst/instruments/niriss/ObservationModes/saii} }
\end{deluxetable}

%%%%%%%%%%%%%%%%%%%%%%%%%%%%%%%%%%%%%%%%%%%%%%%%%%%%%%%%%%%%%%%%%%%%%%%%%%%%%%%%
%
% FIGURES
%

\end{document}